\documentclass[final]{clv2025}
\usepackage{graphicx} 
\usepackage{xcolor}
\usepackage{tipa}
\usepackage{multirow}
\usepackage{comment}
\usepackage{subcaption}
\usepackage{hyperref}
\usepackage{amsmath} 
\usepackage{booktabs} 
\newcommand{\todo}[1]{\textcolor{red}{\textbf{TODO: #1}}}
\usepackage{natbib}
\date{February 2026}

\jvol{vv}
\jnum{nn}
\jyear{2025}

\usepackage{amsmath}
\usepackage{booktabs}

\runningtitle{Articulatory MTL for Non-Canonical Phoneme Recognition}
\runningauthor{Riaz et al.}

\begin{document}

\title{Multi-Task Learning for Non-Canonical Phoneme Recognition via Articulatory Feature Decomposition}

\author{
Sophia Riaz\thanks{Equal contribution}$^{1}$,Haoze Zheng$^{*,1}$, Amos Roche$^{1}$, Miyu Zhang$^{1}$, Anamika Ragu$^{1}$, Salvatore Penachio$^{1}$, Kaustav Mukherjee$^{1}$, Aneesh Jonelagadda\thanks{Corresponding author}$^{1}$}

\affilblock{
    \affil{Kaliber AI, San Mateo CA\\
}}

    \maketitle

\begin{abstract}
Pathological and more broadly non-canonical speech present significant challenges for automatic phoneme recognition due to systematic deviations from canonical pronunciation and limited availability of labeled clinical speech data. Existing phoneme recognition systems are typically trained on canonical speech and treat phonemes as atomic categorical labels, limiting their ability to detect structured articulatory errors common in speech disorders and accents. In this work, we introduce a linguistically structured approach to non-canonical phoneme recognition that decomposes phoneme prediction into articulatory feature dimensions such as manner, place, and voicing. We implement this formulation using a hierarchical multi-task learning architecture in which task-specific articulatory feature heads learn feature-level representations that are subsequently integrated through a cross-attention-based fusion module to produce phoneme predictions. To address the scarcity and noise of pathological speech labels, we combine this framework with semi-supervised learning via Momentum Pseudo-Labeling (MPL) and propose a cascaded training strategy that progressively introduces articulatory feature tasks while employing staged unfreezing of a pretrained speech encoder. Experiments on L2-ARCTIC, used as a proxy for pathological speech variation, show that the proposed approach achieves substantial improvements in phoneme recognition performance compared to strong baseline architectures, while yielding interpretable error patterns aligned with phonological feature structure. These results suggest that articulatory feature supervision is a promising strategy for robust and interpretable phoneme recognition in non-canonical speech, and motivate future validation on clinically diagnosed pathological speech datasets.

\end{abstract}

\maketitle

\section{Introduction}

Spoken language is the primary modality of human communication. Through speech, humans convey not only linguistic content, but also emotion, context, and even indicators of health. When speech production is impaired, often to the extent of a disorder, this communication becomes degraded, reducing the effectiveness of such conveyance. As public awareness of speech disorders has increased, so too has the demand for clinical intervention \cite{asha2023poll}, including at-home speech therapy which entails immediate and detailed feedback on speech production. This requires systems that not only detect incorrect utterances but also provide fine-grained error analysis. Such analysis can be performed at both the phoneme and prosodic levels. While prosodic analysis is relatively robust in modern audio processing pipelines \cite{cohen2025}, phoneme-level analysis remains a significant challenge.

Pathological phoneme recognition has substantial room for improvement. Existing phoneme recognition models are primarily trained on healthy speech, and thus struggle to generalize to disordered speech \cite{berisha2024clinicalai}. Models that incorporate pathological data typically have limited coverage of the spectrum of disorders. As a result, performance on common conditions such as Motor Speech Disorder (MSD) and Childhood Apraxia of Speech (CAS) remains poor, with reported Phoneme Error Rates (PER) ranging from 42\% to 69\% \cite{li2020allosaurus,ravanelli2021speechbrain,baevski2020wav2vec2}. When trained exclusively on healthy speech, these models exhibit a tendency to normalize atypical pronunciations to their closest canonical forms. For example, an utterance of “wapit” (\textipa{/\"w\ae pIt/}) may be incorrectly mapped to the canonical phoneme sequence for “rabbit” (\textipa{/\"r\ae bIt/}). This auto-corrective bias arises because pathological pronunciations are not represented in the training data, highlighting the need for models explicitly trained on non-canonical speech. However, training on pathological speech is fundamentally constrained by the scarcity of high-quality labeled data \cite{torgo2012,uaspeech,nemours} with ground-truth annotations that may themselves be noisy due to inter-annotator variability \cite{strombergsson2020}. Momentum pseudo-labeling (MPL) \cite{yang2022} provides a natural approach to mitigating this through a teacher–student framework that leverages unlabeled data.

Despite advances in self-supervised speech models such as Wav2Vec2 and WavLM \cite{baevski2020wav2vec2,chen2022wavlm}, existing models treat phonemes as atomic categorical labels and do not explicitly encode phonological structure. While effective for canonical speech, this is poorly suited to non-canonical speech where errors are often systematic: speakers may preserve certain articulatory properties while deviating along others \cite{duffy2019}. Articulatory features provide a linguistically motivated decomposition of phonemes that correspond to physical properties of speech production.

To address these challenges, we propose a hierarchical articulatory multi-task learning framework for low-resource non-canonical phoneme recognition that jointly predicts articulatory features and phonemes using cross-attention-based feature fusion. We further incorporate MPL and speech-specific data augmentation to improve robustness. We evaluate our approach on L2-ARCTIC using Phoneme Error Rate (PER), mispronunciation detection metrics, and ablation studies. Beyond overall performance, we analyze error patterns along articulatory feature dimensions. Our results demonstrate consistent improvements in PER over strong baselines and highlight the effectiveness or structured articulatory supervision for low-resource non-canonical settings.

We therefore investigate the following research questions:
\begin{enumerate}
    \item To what extent can articulatory structure be incorporated into phoneme recognition through hierarchical articulatory feature supervision? 
    \item Do phoneme recognition errors exhibit structure along articulatory feature dimensions, and can modeling these dimensions reduce such errors?
    \item Can our novel architecture trained on limited non-canonical speech detect non-canonical speech patterns better than baseline models trained on a much larger corpus of canonical speech?
\end{enumerate}

\subsection{Main Contributions}
Our main contributions are as follows:
\begin{itemize}
\item We introduce a hierarchical multi-task learning (HMTL) architecture, including heuristically-motivated augmentations, for non-canonical phoneme recognition that decomposes phoneme prediction into articulatory feature dimensions integrated through a cross-attention–based fusion layer to produce final phoneme predictions.

\item We propose a structured articulatory supervision framework that derives rich auxiliary labels from ARPAbet and exploits linguistic relationships between phonemes. 

\item We conduct a comprehensive empirical evaluation with detailed ablations, comparisons and analysis where we demonstrate that phoneme recognition errors exhibit systematic structure along articulatory features, motivating the HMTL architecture.
\end{itemize}

\section{Background}
\subsection{Clinical and Linguistic Foundations of Motor Speech Disorders}

Motor speech disorders are characterized by impairments in the processes that transform linguistic intent into coordinated speech output. These processes span multiple stages, from motor planning to the execution of articulatory movements. Two main types prevail in clinical literature: apraxia of speech and dysarthria.

Childhood apraxia of speech (CAS) is a rare neurological speech disorder in which the brain struggles to plan and sequence the complex muscle movements needed for speech despite presenting without neuromuscular deficits. This contributes to a wide range of symptoms, including distorted sounds, inconsistent errors in speech, groping for sounds, and incorrect prosody. These symptoms manifest in a wide array of phonetic errors. Unlike CAS, where motor planning is disrupted, dysarthric speech is the result of impaired motor execution. Six presentations arise from distinct underlying pathologies \cite{clevelandclinic2025dysarthria} and can be broadly characterized in terms of linguistic impairments and alterations in speech quality. The breadth of these deviations from typical speech presents a significant challenge for present automatic speech recognition systems, thereby motivating the need for strategies that better capture the manifestations of this variability.

\subsection{Phoneme Recognition for Non-Canonical Speech}

Fundamentally, the development of Automatic Speech Recognition (ASR) models is geared towards achieving the lowest corpus-level word error rate (WER), underscoring that frontier systems are optimized for word-level transcription accuracy rather than for the preservation of fine-grained phonetic detail \cite{FATEHI2025103151}. The same inductive biases that enable low WER, such as strong language modeling and distributional smoothing, also encourage normalization of disordered speech, mapping atypical pronunciations to the nearest valid phoneme. While this abstraction is beneficial for robust transcription of speech, it obscures the subtle deviations in articulation that are core to diagnosing and treating pathologies arising from motor speech disorders. Transitioning from word-level ASR to phoneme recognition introduces a distinct set of modeling challenges that have important architectural implications. Contemporary ASR pipelines measure mispronunciation through detection of deviations relative to discrete word and subword level targets but they rarely characterize \textit{how} mispronunciations manifest acoustically. At a phoneme level this implementation is not robust enough as clinical descriptions of disordered speech are inherently feature-based, emphasizing dimensions such as articulatory precision, prosody, and voicing control. Additionally, temporal resolution becomes significantly more critical and models must be able to capture fine-grained acoustic transitions corresponding to rapid articulatory movements. Unlike words or subwords, phonemes exhibit high context dependence (e.g. co-articulation effects), shorter durations, and greater acoustic overlap, necessitating architectures that can preserve local temporal detail while still modeling longer-range dependencies.

\subsection{Pathology Data Scarcity and L2 as Proxy}

The challenges of pathological ASR are compounded by the scarcity of high-quality labeled clinical speech data. Collecting and annotating such data is complicated, resulting in datasets that are orders of magnitude smaller than those used in standard ASR, and where inter-annotator variability can be substantial \cite{HOSOM2009352}. This motivates the search for a proxy dataset that can minimally identify mispronunciations in its ground truth. Therefore, we adopt L2-ARCTIC, an accented corpus with aligned phonemic annotations. Although derived from non-native speech, it provides systematic pronunciation deviations while maintaining annotation quality, offering the most practical reconciliation between scale and supervision \cite{l2arctic}. These typologically distinct non-native speakers introduce systematic phoneme variation, where L1 phonological constraints shape L2 articulation in a manner that is structurally comparable to certain pathological deviation patterns \cite{FARISH2020100910,flege,besttyler}. Namely, reduced vowel space, common among East Asian speakers, resembles the centralized vowel production observed in hypokinetic dysarthria \cite{Kent2003-ya}. These accent-driven deviations are predictable within each L1 group, but collectively produce a range of phonetic variability mirroring disordered speech. Nevertheless, while accented speech constitutes the most phonetically comprehensive proxy available, it does not represent the entire spectrum of variability observed in clinical populations, especially impairments caused by respiratory, phonatory, and prosodic systems \cite{duffy2019}. Thus, closing the gap requires strategies that can simultaneously leverage unlabeled data while diversifying the training dataset.


\subsection{Articulatory Feature Representations}
Phoneme production can be decomposed along articulatory dimensions that describe the configuration of the speech apparatus during sound generation. These categories vary between consonants and vowels. Key attributes for consonants include \textbf{place} of articulation, which specifies the location in the vocal tract where air flow is obstructed, \textbf{manner} of articulation which describes how the airflow is modified, and \textbf{voicing} which indicates if the vocal chords are vibrating. Vowels, in contrast, are characterized by tongue position and shape: \textbf{backness}, which represents tongue position, \textbf{height} which reflects the vertical tongue position, and \textbf{roundedness}, indicating lip roundness. Each phoneme can be represented as a combination of such features. Importantly, these features are not all mutually exclusive. Demonstrably, \textipa{/r/} can occupy multiple place categories according to dialect and individual \cite{ladefoged2014course}. Complementing these features, phonological structures operate more abstractly, encoding how phonemes sound within a language rather than their physiological catalysts. Features such as rhoticity encode patterns of linguistic behavior that may not map to an articulatory configuration. Together, phonological and articulatory features form a structured, multi-dimensional representation of speech. This structured representation suggests that phoneme recognition is more naturally formulated as a multi-label prediction problem rather than as classification over independent atomic labels. Furthermore, articulatory features exhibit hierarchical dependencies as vowel and consonant categories constrain the set of valid feature combinations. 

\subsection{Related Work}
The most direct precursor to this work is authored by \citet{yang2022}, who propose a mispronunciation detection framework built on Wav2Vec2.0 and MPL for accented (L2) speech. Their approach demonstrates that combining self-supervised acoustic representations with a teacher—student semi-supervised training regime achieves improvements over purely supervised baselines, motivating our adoption of MPL for low-resource non-canonical speech. However, several limitations remain in the aforementioned work \cite{yang2022}. The formulation remains largely phoneme-centric and as a sequence of atomic phoneme labels. Thus, deviations are reduced to correctness scores without capturing the underlying articulatory structure of errors. Furthermore, robustness to non-canonical acoustic variability relies primarily on the pretrained backbone, whereas our framework incorporates speech-specific data augmentation to improve generalization under diverse pronunciation patterns.

Existing literature explores phonological and articulatory features. Classical approaches model speech as binary compositions of articulatory features instead of discrete labels. Early approaches modeled speech as binary articulatory compositions \cite{stouten2006speech}, while \citet{PanPhonMortensen-et-al:2016} provides linguistically motivated feature inventories. More recent work has revisited these representations with deep learning. Weakly supervised phonological bottlenecks have been used for intelligibility prediction \cite{Thienpondt_2025}, while Phonet estimates frame-level phonological posteriors \cite{tadavarthy2024phonological}. Multilingual speech modeling has benefited from phonological representations \cite{xu2022simple}, which have also been directly supervised for sequence prediction \cite{SHAHIN2025103249}. Despite these advances, articulatory information is primarily employed as an intermediate representation, bottleneck, or downstream feature to improve or analyze speech representations, rather than serving as the central inductive bias for phoneme recognition.

Multi-task learning (MTL) has been widely adopted to improve phoneme recognition and mispronunciation detection by jointly optimizing auxiliary objectives alongside a primary task. \citet{elkheir2023multiview} propose a multi-view multi-task framework that combines multilingual and monolingual encoders with auxiliary prediction of place and manner, improving phoneme recognition in low-resource settings. However, the auxiliary supervision is implemented as parallel classification heads and is limited to a subset of articulatory categories. Similarly, \citet{PANG2026131930} introduce a hierarchical MTL framework for pronunciation assessment, incorporating attention mechanisms to model correlations across linguistic levels. While hierarchical, the supervision remains focused on scoring rather than explicit phoneme-feature decomposition. \citet{Glocker2024HMTL} propose a hierarchical MTL framework for cross-lingual phoneme recognition by conditioning phoneme prediction on articulatory attribute predictions. Their work focuses on multilingual transfer using binary articulatory attributes and propagates probability distributions to the phoneme classifier. In contrast, we supervise a set of articulatory categories, integrating learned representations through cross-attention prior to phoneme decoding. 

Collectively, these demonstrate that structured auxiliary supervision to improve learning is promising. However, current approaches either employ parallel supervision, target pronunciation or multilingual transfer, or condition prediction on non-articulatory features or probabilities. There remains limited work on hierarchical articulatory supervision for non-canonical phoneme recognition, grounded in the underlying physiology of speech production in low-resource conditions.

\section{Methods}

\begin{figure}[b]
    \centering
        \includegraphics[width=\textwidth]{pch.jpg}
    \caption{Comparison of the IPA consonant inventory and the derived ARPAbet articulatory inventory used in this work. The ARPAbet inventory was constructed from the L2-ARCTIC ARPAbet–IPA phoneme correspondence and organized according to IPA articulatory feature categories.}
    \label{fig:arpa_ipa_cons_chart_final}
\end{figure}

\subsection{Articulatory Feature Derivation}

WavLM is pre-trained on the Librispeech corpus \cite{librispeech}, and further fine-tuned using L2-ARCTIC, which both primarily adopt ARPAbet phonemic transcriptions \cite{l2arctic}. While ARPAbet provides a compact representation of speech, standard articulatory feature mappings are typically defined over IPA \cite{ipa,arpabet,ladefoged2014course}. Consequently, the phoneme labels available for training do not explicitly encode the articulatory structure underlying speech production. This is exemplified by L2-ARCTIC's selective phoneme inventory, further reducing phonetic granularity from 89 standard IPA phonemes excluding diacritics \cite{ipa} to a reduced set of 40 phonemes. Notably, there is an absence of transcribed non-English articulations in the current constrained ARPAbet space despite being present in the corpus' L1 backgrounds. As a result, the resulting phonemic representations fail to capture the full richness and cross-linguistic variation essential for modeling accented or pathological speech.


To obtain articulatory features, each ARPAbet phoneme was mapped to a set of linguistically motivated articulatory attributes. IPA representations' articulatory mappings were used an as intermediate resource to define these. By decomposing ARPAbet phonemes into articulatory properties, the resulting feature representation allows the model to detect the corresponding ARPAbet phoneme while also being able to predict the particular features associated with it. Figure \ref{fig:arpa_ipa_cons_chart_final} illustrates how the reduced ARPAbet inventory collapses articulatory distinctions, leading to ambiguity and motivating the use of auxiliary articulatory supervision. The modeling process is further complicated by the existence of diphthongs and affricates. ARPAbet encodes these as single symbolic units, whereas IPA explicitly encodes their composite structure. This obscures internal phonetic structure and limits the model's ability to learn fine-grained temporal and articulatory transitions. In addition to this, phonemic transcriptions for both ARPA and IPA omit phonetic detail conveyed through diacritics and suprasegmental markers. We exhaustively enumerate these, as well as the absent non-native transcriptions, in the Appendix (Table \ref{tab:lang_arpa_ipa_missing}, Table \ref{tab:arpa_ipa_dipthong}).

Taken together, these limitations highlight a critical gap between the phonetic richness required for accurate pronunciation modeling and the constrained symbolic representations used in current speech datasets. To address these limitations, we derive articulatory feature labels from ARPAbet phonemes, enabling the model to learn linguistically meaningful intermediate representations while maintaining compatibility with the corpus's phoneme inventory.

\subsection{Articulatory Feature Multi-Task Learning Architecture}

\begin{figure}[ht] 
    \centering
    \includegraphics[width=0.85\linewidth]{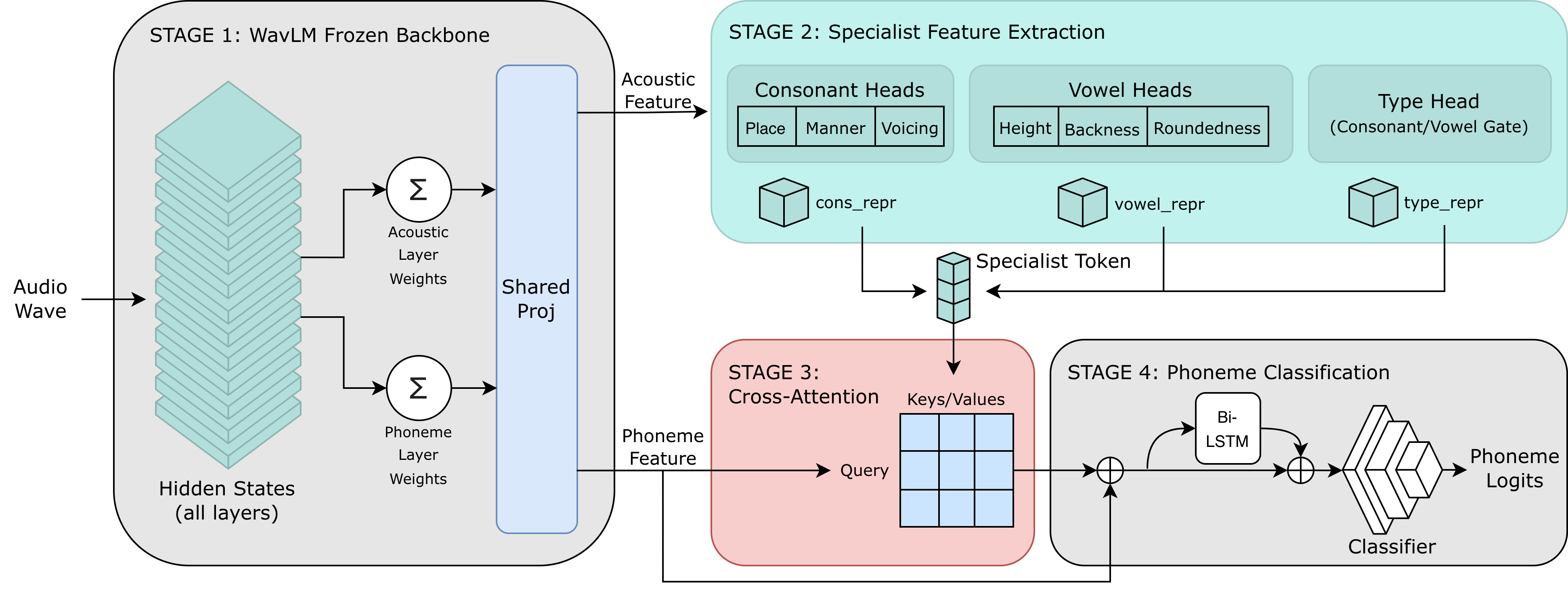}
    \caption{Hierarchical MTL architecture. (Stage 1) Input audio is first passed into a frozen WavLM backbone, where acoustic and phoneme representations are learned as two separate weighted sums of transformer layers. (Stage 2) Acoustic features are passed into separate articulatory feature estimation heads for consonant, vowel, and type representations. (Stage 3) These, along with the phoneme WavLM features are passed into a cross-attention layer, whose output is residually fused with the phoneme features. (Stage 4) This is then passed through a bidirectional LSTM to capture temporal dependencies before the final phoneme classification head.    
    }
    \label{fig:hmtl}
\end{figure}


We introduce a hierarchical multi-task learning (HMTL) architecture for frame-level recognition from raw speech (Figure \ref{fig:hmtl}). The model decomposes phoneme prediction into several stages which can be grouped into first articulatory feature estimation, then phoneme classification, which is strengthened by the articulatory features. The architecture given an input waveform produces predictions for auxiliary articulatory features such as phoneme type, place, manner, voicing, height, backness, and roundedness to use those logits to strengthen a phoneme classification model.

\textbf{Stage 1: Acoustic Feature Extraction with Learnable Shared Projections}

We use a pretrained WavLM \cite{chen2022wavlm} model as the acoustic backbone to extract frame-level representation from raw waveform inputs. The model outputs hidden states from all of the transformer layers. All backbone parameters are initially frozen, allowing the auxiliary heads to learn an initial state of information utilizing a higher learning rate. Instead of relying on a single hidden layer from WavLM, we learn two combinations of layer representations. The acoustic representation is used primarily for articulatory prediction tasks and the phoneme representation is used for the final phoneme classification. Both representations are computed as a learnable weighted sum of hidden states $h$ over $L$ total WavLM layers:

\[
\tilde{h} = \sum_{l=1}^{L} \alpha_l h^{(l)}, \quad \alpha = \text{softmax}(w)
\]

where $w \in \mathbb{R}^L$ is a learnable parameter vector and $\alpha$ defines normalized layer weights. Both the acoustic and phoneme representations are then passed through a shared projection block, consisting of a linear transformation, layer normalization, GELU activation function, and dropout layer for training regularization.

\textbf{Stage 2: Articulatory Specialist Heads}

To model the phonological structure, the learned acoustic representation is passed into three articulatory specialist branches: a \textbf{phoneme type} branch, a \textbf{consonant} branch, and a \textbf{vowel} branch. These branches produce task-specific latent representations that are later used to guide phoneme classification. Each prediction head in the model follows a two-layer feedforward architecture. Specifically, each head consists of a linear projection from the input dimension to a hidden dimension, followed by a layer normalization, GELU activation, a dropout layer, and then a final linear layer that maps to the task output. This design provides a lightweight but expressive prediction module for each auxiliary task. The type branch is a binary classification head which predicts whether each frame corresponds to a consonant or vowel. This representation serves as a gating mechanism, allowing the model to distinguish between a vowel and consonant to strengthen the processing of the consonant/vowel articulatory heads. 

Together, these three branches form a set of articulatory features that decompose speech into meaningful components before phoneme classification. By modeling articulatory structure, the model provides intermediate representations that can be leveraged in the downstream phoneme prediction stage. 

\textbf{Stage 3: Specialist Memory Cross-Attention}
To integrate articulatory knowledge into the phoneme prediction, we introduce a cross-attention mechanism that enables phoneme branch to dynamically attend to the specialized articulatory representations at each time step. This is performed by first introducing temporal concatenation as a structure in the architecture.
Let $\mathbf{h}^{(p)}_t \in \mathbb{R}^{d}$ denote the phoneme-branch representation at frame $t$, and let $\mathbf{h}^{(c)}_t$, $\mathbf{h}^{(v)}_t$, $\mathbf{h}^{(y)}_t \in \mathbb{R}^{d}$ denote the consonant, vowel, and type specialist representations, respectively. These are stacked into a specialist memory matrix:
\[
\mathbf{S}_t = \begin{bmatrix} \mathbf{h}^{(c)}_t \\ \mathbf{h}^{(v)}_t \\ \mathbf{h}^{(y)}_t \end{bmatrix} \in \mathbb{R}^{3 \times d}.
\]
The phoneme representation $\mathbf{h}^{(p)}_t$ serves as the query, representing the model's current estimate of phoneme-relevant acoustic context before incorporating articulatory information. The specialist memory is projected into key and value spaces via learnable matrices $W_K, W_V \in \mathbb{R}^{d \times d_k}$:
\[
Q_t = \mathbf{h}_t^{(p)}, \quad K_t = \mathbf{S}_t W_K, \quad V_t = \mathbf{S}_t W_V, \quad K_t, V_t \in \mathbb{R}^{3 \times d_k}.
\]
The cross-attention output is then computed as:
\[
\mathbf{c}_t = \operatorname{softmax}\!\left( \frac{\mathbf{h}^{(p)}_t \, K_t^\top}{\sqrt{d_k}} \right) V_t,
\]
where the softmax produces normalized attention weights $\alpha_{t,i}$ over the three specialist tokens, and $\mathbf{c}_t = \sum_{i=1}^{3} \alpha_{t,i} \, \mathbf{v}_{t,i}$ is the resulting articulatory context vector. The scaling factor $1/\sqrt{d_k}$ stabilizes training by controlling the magnitude of the dot-product scores. 

In contrast to simple concatenation, which treats all articulatory feature representations as equally informative, this cross-attention mechanism allows the learned phoneme representation to selectively attend to feature-specific signals at each time step. This is particularly important for resolving phoneme ambiguity, where the importance of articulatory dimensions (e.g., voicing, place, or manner) varies depending on the acoustic context. By dynamically weighting these feature representations, the model can more effectively disambiguate phonemes that are acoustically similar but differ along specific articulatory dimensions.

\textbf{Stage 4: Fusion and Output branch}

The articulatory context vector is fused with the phoneme representation via a residual connection:
\[
\tilde{\mathbf{h}}_t = \operatorname{LayerNorm}\!\left(\mathbf{h}^{(p)}_t + \mathbf{c}_t\right).
\]
The residual formulation preserves the original phoneme encoding while augmenting it with articulatory evidence. Layer normalization stabilizes training by reconciling the feature distributions of the two components, which originate from different latent subspaces. The fused representation $\tilde{\mathbf{h}}_t$ is passed through a bidirectional LSTM \cite{bilstmgraves} to capture temporal dependencies:
\[
\mathbf{h}^{\mathrm{temporal}}_t = \operatorname{BiLSTM}(\tilde{\mathbf{h}}_t) = \left[ \overrightarrow{\mathbf{h}}_t \; ; \; \overleftarrow{\mathbf{h}}_t \right],
\]
where $\overrightarrow{\mathbf{h}}_t$ and $\overleftarrow{\mathbf{h}}_t$ encode past and future context, respectively. Because phonemes are influenced by their neighbors, frame-level predictions based on local features alone are often insufficient. The BiLSTM models contextual dependencies across time, smoothing noisy frame-level representations and propagating the attended articulatory features across the sequence. Finally, the temporal representation is passed to a final classification head:
\[
\mathbf{y}_t = f_{\mathrm{phoneme}}\!\left(\mathbf{h}^{\mathrm{temporal}}_t\right),
\]
producing frame-level phoneme logits for Connectionist Temporal Classification (CTC) decoding \cite{ctcgraves}.

\textbf{Multi-Task Learning Objective}
To jointly optimize phoneme recognition and articulatory feature prediction, we adopt a MTL framework with adaptive task weighting. Rather than relying on fixed loss weights, we employ a Pareto-based weighting that dynamically balances task objectives during training \cite{ozansener}.

The primary phoneme task is trained with CTC loss: 
\[
\mathcal{L}_{\mathrm{phoneme}} = \mathrm{CTC}\!\left(\mathbf{y}^{(\mathrm{phoneme})}, \hat{\mathbf{y}}^{(\mathrm{phoneme})}\right),
\]

where $\hat{\mathbf{y}}^{(\mathrm{phoneme})}$ denotes the frame-level phoneme logits and $\mathbf{y}^{(\mathrm{phoneme})}$ the ground-truth phoneme sequence. The auxiliary articulatory tasks use frame-level cross-entropy for multi-class targets (type, place, manner, height, and backness) and binary cross-entropy for binary targets (voicing and roundedness), each computed only over frames belonging to the relevant phoneme category (consonant or vowel).

The phoneme loss is assigned a fixed weight of $1.0$, while the auxiliary task weights are determined via Multiple-Gradient Descent Algorithm (MGDA) \cite{mgda}. Let $\mathbf{g}_i = \nabla_{\boldsymbol{\theta}_s} \mathcal{L}_i$ denote the gradient of the $i$-th auxiliary loss with respect to the shared representation $\boldsymbol{\theta}_s$. MGDA finds Pareto-optimal weights $\boldsymbol{\alpha} = (\alpha_1, \ldots, \alpha_K)$ by solving:
\[
\min_{\boldsymbol{\alpha} \in \Delta^K} \left\lVert \sum_{i=1}^{K} \alpha_i \, \hat{\mathbf{g}}_i \right\rVert^2,
\]
where $\Delta^K$ is the probability simplex and $\hat{\mathbf{g}}_i = \mathbf{g}_i / \lVert \mathbf{g}_i \rVert$ are the $\ell_2$-normalized task gradients. This quadratic program is solved efficiently via the Frank--Wolfe algorithm at each training step.
The overall training objective is:
\[
\mathcal{L} = \mathcal{L}_{\mathrm{phoneme}} + \lambda_s \sum_{i=1}^{K} \alpha_i \, \mathcal{L}_i,
\]
where $\lambda_s$ is dependent on the stage and is adjusted throughout the training phases to progressively focus on the primary phoneme task.
By decoupling the phoneme loss from the Pareto optimization, we ensure that auxiliary articulatory tasks inform the shared representation without harming phoneme convergence. The MGDA weights prevent any single auxiliary task from dominating the gradient, promoting balanced learning across the articulatory feature space.



\textbf{Training Strategy}

To ensure stable optimization and effective learning of both acoustic and articulatory representations, we adopt a progressive training strategy consisting of two stages. In the first stage, the pretrained WavLM backbone is fully frozen, and only the newly introduced components are trained. During this phase, we use a relatively higher learning rate for the trainable parameters. This allows the newly initialized layers to adapt to downstream phoneme recognition tasks and learn meaningful representations without interfering with pretrained acoustic features. Since the WavLM representation starts off as fixed, the heads learn a stable mapping without gradient interference. In the second stage, we unfreeze the top k layers of the WavLM encoder and jointly optimize both the backbone and head parameters. Unfreezing only the higher layers of WavLM allows the model to refine higher-level acoustic representations while preserving low-level features during the pretraining. To prevent catastrophic forgetting and overfitting, we adopt a smaller learning rate for both the backbone and head parameters. In addition, we incorporate a replay mechanism in which a subset of previously seen samples are periodically revisited during training to further mitigate catastrophic forgetting and stabilize the fine-tuning process. 
Overall, the training procedure can be interpreted as a two-phase optimization problem: an initial representation alignment phase, followed by a joint refinement phase. The first phase learns a stable mapping from pretrained features, while the second phase refines both the head and higher layer WavLM features jointly. 

\subsection{Momentum-Based Pseudo Labeling}
As discussed previously, phoneme labels are often noisy and limited. These challenges are further intensified in pathological and non-canonical speech, where rarer conditions may have little to no labeled data available. To improve generalization in such settings and leverage additional unlabeled speech, we extend our multi-task framework with a semi-supervised momentum-based learning strategy called momentum pseudo labeling. This approach integrates labeled data with a teacher-student distillation setup applied to unlabeled inputs, enabling the model to learn more robust representations.

\textbf{Data Setup}

1. Labeled dataset $\mathcal{D}_L = \{(x_i, y_i)\}$:
    Provides phoneme sequences and articulatory feature annotations for all tasks.
    
2. Unlabeled dataset $\mathcal{D}_U = \{x_j\}$:
    Contains raw speech waveforms without annotations, used for consistency-based learning.

For unlabeled samples, a clean waveform x is used by the teacher model and an augmented waveform x is used by the student model.

\textbf{Momentum Teacher}

We maintain a teacher model $f_{\theta'}$ whose parameters are updated as an exponential moving average (EMA) of the student model $f_{\theta}$:
\[
\theta' \leftarrow \alpha \theta' + (1 - \alpha)\theta
\]

where $\alpha \in [0,1)$ controls the update rate.

This produces a temporally smoothed model that serves as a source of pseudo-labels for unlabeled data. 

\textbf{Learning on Unlabeled Data}

For unlabeled inputs $x \sim \mathcal{D}_U$, the teacher produces soft pseudo-labels from the clean signal:
\[
p_{\theta'}(y \mid x)
\]

The student processes an augmented version $x^{\mathrm{aug}}$ and produces:
\[
p_{\theta}(y \mid x^{\mathrm{aug}})
\]

Rather than using strict pseudo-labels, we enforce consistency between the two distributions:

\[
\mathcal{L}_{\mathrm{unsup}} = \mathbb{E}_{x \sim \mathcal{D}_U} \left[
D\left( p_{\theta}(y \mid x^{\mathrm{aug}}), \, p_{\theta'}(y \mid x) \right)
\right]
\]

The total unsupervised loss consists of a KL divergence for all of the categorical heads:

\begin{equation}
\mathcal{L}_{\mathrm{cat}}^{\mathrm{unsup}} = \mathrm{KL}\left(p_{\theta'}  p_{\theta}\right)
\end{equation}

The L2 distance is used for the binary heads:

\begin{align}
    \mathcal{L}_{\mathrm{bin}}^{\mathrm{unsup}} =  \sigma(z_{\theta'}) - \sigma(z_{\theta}) ^2
\end{align}

The loss is then aggregated as a sum. The complete objective combines the total supervised loss and unsupervised loss. 

Although some explicit frame-level annotations are unavailable, the teacher produces a sequence of probability distributions over phoneme tokens at each time step. These distributions are soft signals allowing the student model to learn where phonemes occur without requiring explicit labels. By minimizing the KL divergence between the teacher and student outputs, the model is able to produce more stable predictions even with noisy data.

\subsection{Speech-based Data Augmentations}
To improve robustness under limited supervision, we apply waveform-level speech augmentations that increase acoustic variability. For an input waveform $x$, augmentation is applied with probability $p_{\mathrm{aug}}$:
\begin{equation}
\tilde{x} =
\begin{cases}
\mathcal{A}(x; \theta), & \text{with probability } p_{\mathrm{aug}},\\
x, & \text{otherwise},
\end{cases}
\end{equation}
where $\mathcal{A}$ is one augmentation operator and $\theta$ denotes its sampled parameters. The available operations are phase perturbation \cite{chengxilei}, VTLP \cite{jaitly2013vocal}, pitch shifting, prosodic time-stretching, and optional additive noise. Their strengths are sampled from predefined ranges in the configuration. Time and frequency masking are implicitly provided by the WavLM frontend through SpecAugment \cite{park19e_interspeech}, and are therefore not re-implemented.

\paragraph{Phase perturbation}
Phase perturbation is performed in the STFT domain. Given
\begin{equation}
X(\omega,\tau)=|X(\omega,\tau)|e^{j\phi(\omega,\tau)},
\end{equation}
we preserve the magnitude and perturb the phase:
\begin{equation}
\tilde{X}(\omega,\tau)=|X(\omega,\tau)|e^{j(\phi(\omega,\tau)+\Delta\phi(\omega,\tau))}.
\end{equation}
The waveform is then reconstructed by inverse STFT. Our code supports standard, frequency-dependent, temporally smoothed, and selective-band variants, with perturbation strength sampled from a configurable interval (default $[0.05,0.25]$). This augmentation is intended to mimic unstable or effortful phonation, as observed in spastic dysarthria.

\paragraph{Vocal Tract Length Perturbation (VTLP)}
VTLP warps the frequency axis,
\begin{equation}
f' = g_{\alpha}(f),
\end{equation}
where $\alpha$ is sampled from a configured warp range. In our implementation, VTLP is applied over the full signal with coverage $1.0$ and cutoff frequency $f_{\mathrm{hi}}=4800$ Hz. This transformation increases variability in the spectral envelope and formant structure, which is relevant to resonance-related deviations such as breathy or dysphonic speech, while also improving invariance to speaker-dependent spectral variation.

\paragraph{Pitch and prosodic perturbation}
Pitch shift modifies the perceived fundamental frequency by a sampled number of semitone steps $n$, while prosodic augmentation applies time-stretching with rate $\gamma$:
\begin{equation}
\tilde{x} = \mathcal{P}(x;n), \qquad \tilde{x}(t)=x(\gamma t).
\end{equation}
The values of $n$ and $\gamma$ are sampled from configured ranges. These augmentations target abnormalities in intonation, speech rate, and stress patterning, and are motivated by disorders such as ataxic, hypokinetic, and hyperkinetic dysarthria, as well as CAS, where prosodic control is frequently disrupted.

\paragraph{Additive noise}
When enabled, noise is added using an externally provided noise source with signal-to-noise ratio sampled from configured bounds. Although additive noise is not a physiological model of disordered speech, it can approximate degraded or breathy voice quality and improves robustness to low-quality acoustic realizations.

Overall, these augmentations should be interpreted as \emph{acoustic proxies} rather than faithful clinical simulations. They do not reproduce the neuromotor mechanisms or coordinated error patterns of real pathological speech, but instead expand the training distribution along clinically relevant dimensions of spectral, phonatory, and prosodic variability.

\section{Experimental Setup}
\subsection{Dataset description.}

We train and evaluate on the L2-ARCTIC corpus, which contains read speech from 24 non-native English speakers spanning six first-language (L1) backgrounds: Vietnamese, Korean, Mandarin, Spanish, Hindi, and Arabic, with four speakers per L1 (balanced for gender)\cite{l2arctic}. Each utterance is accompanied by Praat TextGrid \cite{praat} annotations providing both \emph{canonical} phoneme transcriptions (the expected pronunciation) and \emph{perceived} phoneme transcriptions (what expert annotators actually heard), along with time-aligned segment boundaries. As previously stated and as supported by previous studies \cite{FARISH2020100910,flege,besttyler}, while this does not explicitly include pathological data, we use the L2 speech as a proxy for pathological speech, mainly noting similarities to dysarthritic speech. 

\textbf{Ground-Truth Establishment.}

Defining ground-truth phoneme labels for non-native speech is inherently challenging: phoneme productions by L2 speakers often fall between canonical categories, and annotator judgments can vary considerably. Prior work on non-native speech transcription reports that inter-annotator agreement at the phone level typically reaches only fair-to-moderate levels, with Cohen's Kappa values often in the range of $0.7$--$0.85$ \cite{ryu12_interspeech}, depending on annotation constraints and task design. The L2-ARCTIC annotations were produced by trained linguists experienced in transcribing non-native speech, with automated consistency checks and secondary verification applied to improve quality. Nevertheless, the inherent subjectivity of labeling accented and mispronounced speech means that ground-truth labels are noisy by nature---a key motivation for our use of MPL, which can leverage unlabeled data to reduce dependence on potentially inconsistent annotations.

\textbf{Preprocessing.}
We apply normalization to the raw annotations before training:
\begin{enumerate}
    \item \textbf{Stress removal.} ARPAbet symbols include lexical stress markers (e.g., \texttt{AH0}, \texttt{AH1}, \texttt{AH2}). We strip all stress digits, collapsing stressed variants into a single phoneme identity (e.g., \texttt{ah}). This reduces the label space without discarding segmental information, since stress distinctions are not the focus of articulatory assessment.
    \item \textbf{Silence normalization.} Silence tokens (\texttt{sil}, \texttt{sp}, \texttt{spn}, \texttt{pau}) are unified under a single \texttt{sil} label. Artificial silences introduced by the L2-ARCTIC annotation scheme are removed, and consecutive segments are collapsed into a single token to prevent the model from over-representing pauses.
    \item \textbf{Phoneme inventory.} After normalization, the inventory comprises 39 non-silence phonemes plus a silence token. For CTC training, silence frames are excluded from the target sequence, yielding a 39-phoneme vocabulary with an appended CTC blank, for a total of 40 output classes.
\end{enumerate}

\textbf{Articulatory Feature Decomposition.}

The seven auxiliary targets are derived deterministically from our ARPAbet-to-articulatory-feature mapping and are computed at the frame level using the time-aligned segment boundaries from the TextGrid annotations. Consonant-specific features (place, manner, voicing) are masked on vowel frames and vice versa, so each auxiliary head is supervised only on phonologically relevant frames.

\textbf{Speaker Splits.}
We partition speakers into 18 for training and 6 for testing, ensuring no speaker overlap between sets. The test speakers represent one speaker from each of the six L1 backgrounds. The training set is further split 90/10 by utterance for training and validation.
\subsection{Experiments overview}

\textbf{Ablation Study}

\textbf{MTL Heads}
To systematically evaluate the contribution of each articulatory auxiliary task to phoneme recognition, we conduct a Leave-One-Out (LOO) ablation study within the proposed multi-task learning (MTL) framework discussed in section 3.2. This analysis is designed to isolate the effect of individual articulatory features. 

In every iteration one articulatory feature head is dropped, following the flow:

Let \( \mathcal{T} \) denote the set of active tasks and \( f_{\theta}^{\mathcal{T}} \) the corresponding model.

The full task set is:
\[
\mathcal{T}_{\text{full}} =
\{\text{phoneme}, \text{type}, \text{place}, \text{manner}, \text{voicing}, \text{height}, \text{backness}, \text{roundedness}\}.
\]

For each auxiliary task \( t \in \mathcal{T}_{\text{aux}} \), we define an ablated model:
\[
\mathcal{T}^{(-t)} = \mathcal{T}_{\text{full}} \setminus \{t\}, \quad
f_{\theta}^{(-t)} = f_{\theta}^{\mathcal{T}^{(-t)}}.
\]

Each model \( f_{\theta}^{(-t)} \) is trained under identical conditions, and performance differences are used to quantify the contribution of task \( t \).

Finally, rather than adopting a conventional soft ablation approach, where task losses are masked while retaining the full model architecture, we employ a hard ablation strategy that modifies the network structure. Specifically, when an articulatory task is removed, its corresponding prediction head is eliminated entirely from the model. 

All ablation experiments are conducted under a simplified training regime, where data augmentations and momentum pseudo-labeling are disabled, and both Stage 1 (frozen backbone) and Stage 2 (progressive unfreezing) are limited to 5 epochs each. Although this setup does not fully converge the model, it provides a consistent basis for comparing architectural variants. As the ablation study focuses on relative differences between configurations rather than absolute performance, this regime is sufficient to capture the contribution of individual components.

\textbf{Data Augmentation}
To assess the contribution of each waveform-level augmentation described in Section 3.4, we conduct a Leave-One-Out (LOO) ablation study over the augmentation pipeline. Starting from the full augmentation setting, which includes phase perturbation, VTLP, pitch shifting, prosodic time-stretching, and optional additive noise, we remove one augmentation at a time while keeping all others unchanged. Each model is trained under the same architecture, optimization settings, and data split, such that performance differences can be attributed to the removed augmentation. Unlike the MTL head ablation, this analysis does not modify the network structure, but only the stochastic training-time transformation pipeline. The purpose of this experiment is to quantify the extent to which each augmentation contributes to robustness by increasing acoustic variability along spectral, phonatory, and prosodic dimensions.

\subsection{Implementation Details}

\begin{table}[!b]
\caption{Model performance on L2-ARCTIC (Phoneme Error Rate, $\%$)}
\centering
  \label{tab:per_results}
\begin{tabular}{p{0.42\textwidth}rrrr}
\toprule
\textbf{Model} &  \textbf{PER} $\downarrow$ & \textbf{Prec} & \textbf{Rec} & \textbf{F1} \\
\midrule
\textbf{Canonical Baselines} \\
Wav2Vec2Phoneme         & 22.76 & 0.311 & 0.368 & 0.337 \\
WavLM-base-plus             & 28.02 & 0.281 & 0.477 & 0.353 \\
\midrule
\multicolumn{5}{l}{\textbf{Fine-Tuned Baselines (WavLM-base-plus)}}\\
Wav2Vec2Phoneme     &  14.81 & 0.540 & 0.479 & 0.507  \\
WavLM  & 14.71 & 0.545 & 0.508 & 0.526  \\
WavLM + Aug                      & 14.72 & 0.539 & 0.467 & 0.500 \\
WavLM + MPL                      & 14.54 & 0.551 & 0.483 & 0.515 \\
WavLM + Aug + MPL                & 14.46 & 0.548 & 0.462 & 0.502 \\
\midrule
\multicolumn{5}{l}{\textbf{Multi-Task Learning (WavLM-base-plus backbone)}}\\
Parallel MTL                       & 14.82 & 0.532 & 0.449 & 0.487 \\
Parallel MTL + Aug.                & 14.55 & 0.544 & 0.462 & 0.500 \\
Parallel MTL + Aug. + MPL          & 14.46 & 0.546 & 0.459 & 0.499 \\
\hline
Hierarchical MTL  (HMTL)                 &    14.23 & 0.554 & 0.534 & 0.544 \\
HMTL + Aug.            &    13.89 & 0.570 & 0.488 & 0.526 \\
HMTL + Aug. + MPL      &    13.68 & 0.576 & 0.514 & 0.543 \\
\hline
HMTL + Cross-attention (CA)              &          13.60 & 0.584 & 0.519 & 0.549 \\
HMTL + CA + Aug.       &          13.52 & 0.587 & 0.511 & 0.546 \\
HMTL + CA + Aug. + MPL &  \textbf{13.47}& 0.586 & 0.524 & 0.553 \\
\hline
\textbf{External Baselines} \\
$MV_{multi} - MT_{seq}$ \cite{elkheir2023multiview} & 14.13 & 0.614 & 0.592 & 0.603\\
\bottomrule
\end{tabular}
\end{table}

The model was implemented in PyTorch with HuggingFace Transformers, using \texttt{microsoft/wavlm-base-plus} as the acoustic encoder. Unlike ASR-fine-tuned encoders, \texttt{wavlm-base-plus} is used here as a general-purpose pretrained speech encoder and does not encode an explicit audio-text mapping before task-specific fine-tuning. The Base-Plus variant was pretrained on 94k hours of 16 kHz speech drawn from Libri-Light, GigaSpeech, and English VoxPopuli, providing substantially broader acoustic coverage than standard base-scale SSL speech models trained on smaller corpora. Training was conducted on NVIDIA RTX A4000, RTX 3090, and GV100 GPUs.

We used a cascaded multi-task architecture with hidden dimension 384 and dropout 0.35. Training proceeded in two stages. In Stage 1, the WavLM backbone was frozen and only the task-specific layers were trained for 15 epochs. In Stage 2, the last 6 WavLM transformer layers were unfrozen and jointly optimized with the task heads for 15 additional epochs. Full-backbone fine-tuning was not used in the final setting.

Optimization used AdamW with separate parameter groups for the task heads and WavLM backbone. The learning rate was $2.00\times10^{-3}$ in Stage 1 for the task heads, and $9.78\times10^{-5}$ / $9.58\times10^{-5}$ for the task heads / WavLM backbone in Stage 2. A cosine annealing scheduler was applied in Stage 2, and gradient clipping with maximum norm 0.35 was used throughout. We trained the model with an MGDA-based multi-task objective and CTC phoneme supervision (blank index 40). PER was computed by greedy CTC decoding followed by Levenshtein distance against the reference phoneme sequence, and validation PER was used for model selection. Hyperparameters were tuned with Optuna using a TPE sampler and MedianPruner. The search included hidden dimension, dropout, batch size, stage-wise learning rates, stage lengths, and the number of unfrozen WavLM layers. The final configuration was selected by minimizing validation PER.

\section{Results and Discussion}
\subsection{Model Evaluation and Comparisons}
\begin{table}[b]
\caption{Changes in phoneme error types after applying MPL.}
\centering
\small
\begin{tabular}{llrrr}
\toprule
Model & Error Type & Before & With MPL & $\Delta$ (\%) \\
\midrule
\multirow{4}{*}{Baseline}
& Substitutions &  3055 & 	3006& 	-1.60  \\
& Deletions     &  673	&   705	&      4.76 \\
& Insertions    &  555	&   495	&      -10.81 \\
& Cross-class   &  123	&   117	&     -4.88 \\
\midrule
\multirow{4}{*}{HMTL}
& Substitutions &  2860	& 2793	& -2.34 \\
& Deletions     &  457	& 534	& 16.85 \\
& Insertions    &  542	& 478	& -11.8 \\
& Cross-class   &   179	& 173	& -3.35 \\
\midrule
\multirow{4}{*}{HMTL with cross-attention}
& Substitutions & 2699	& 2732	& 1.22 \\
& Deletions     &  576	& 520	& -9.72 \\
& Insertions    &  488	& 501	& 2.66 \\
& Cross-class   &  169	& 165  	& -2.37 \\
\bottomrule
\end{tabular}
\label{tab:mpl_error_analysis}
\end{table}

We evaluate multiple model variants: baseline phoneme recognition models such as Wav2Vec and WavLM, augmented baselines, and our proposed multi-task learning (MTL) architectures (Table \ref{tab:per_results}), which outperforms the multi-view sequential MTL framework proposed by \citet{elkheir2023multiview}. Performance is measured using phoneme error rate (PER). We see that zero-shot performance (where we do not train on a subset of the L2-Arctic dataset) is significantly lower than all of the non-canonical fine-tuned models. This addresses our first research question and paints a strong motivation for fine-tuning on a non-canonical speech dataset. Among fine-tuned baseline systems, WavLM slightly outperforms Wav2Vec2, likely due to its utterance mixing strategy and gated relative positional bias, which better capture temporal dependencies in speech. Incorporating data augmentation consistently improves performance, suggesting that exposure to greater acoustic variability improves model robustness, particularly for unstable or low-energy phonetic segments. Incorporating MPL yields further gains across model variants. While MPL consistently improves overall PER, its effect on individual error types depends on the underlying architecture. As shown in Table \ref{tab:mpl_error_analysis}, MPL reduces insertion and substitution errors for both the baseline and HMTL model ($-10.8\%$ and $-11.8\%$, respectively), although this is accompanied by increased deletions ($+4.8\%$ and $+16.9\%$), suggesting a more conservative decoding strategy. In contrast, when cross-attention is introduced into HMTL, MPL instead reduces deletions ($-9.7\%$) while slightly increasing insertions and substitutions. Despite these differing error profiles, overall PER improves, indicating that MPL primarily enhances generalization and sequence alignment, with its influence on insertion-deletion trade-offs depending on the underlying architecture. Introducing MTL in a parallel formulation yields marginal improvements over the baseline, suggesting that naively incorporating auxiliary objectives does not improve performance and may instead introduce optimization challenges. A more substantial improvement is observed with hierarchical MTL architectures and cross-attention, which explicitly structure the learning of articulatory features by progressively supervising higher-level representations. These results indicate that performance gains are not driven by any single component, but by the interaction between model architecture, linguistic structure, data augmentation, and semi-supervised learning. While augmentation and MPL enhance generalization, the largest improvements arise when phoneme prediction is built with linguistically motivated auxiliary tasks. This motivates the subsequent analysis of error patterns to better understand where these improvements originate.

\paragraph{Best Baseline Model Error Analysis}

\begin{figure}[t]
    \centering
        \includegraphics[width=\textwidth]{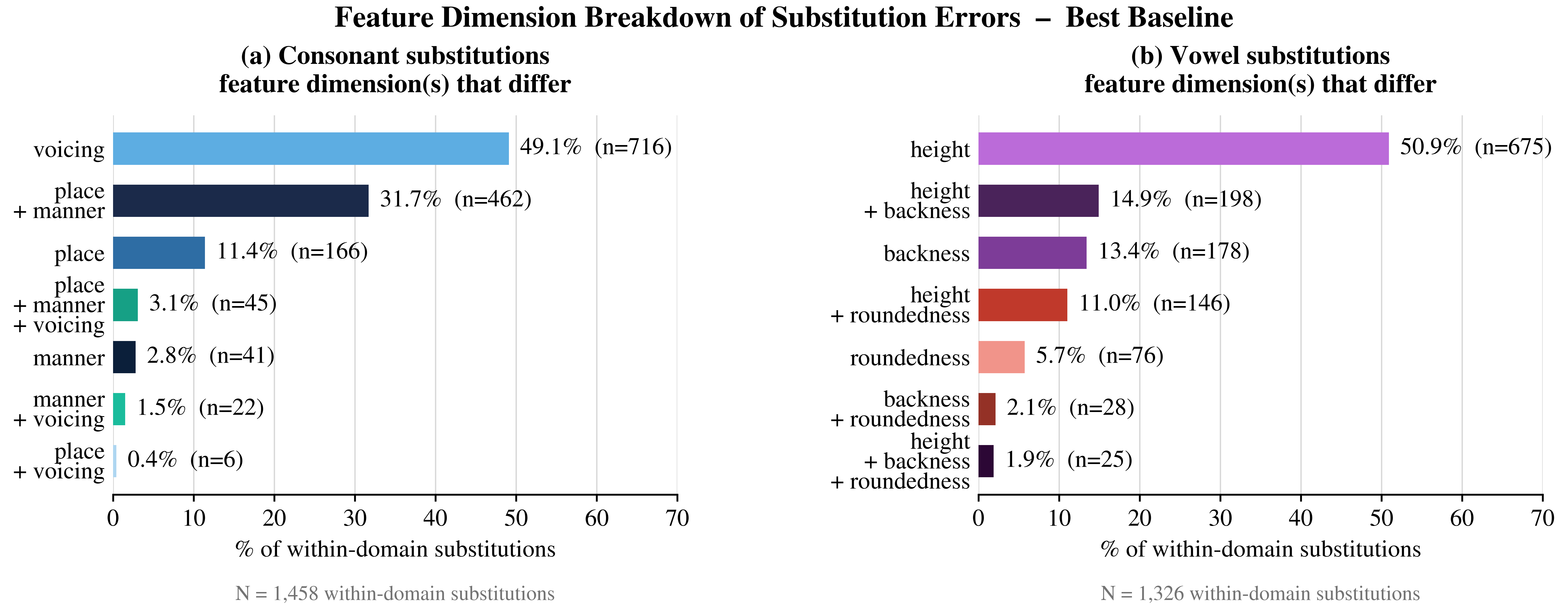}
    \caption{Breakdown of within-domain substitution errors by articulatory feature dimensions for the best baseline model.}
    \label{fig:error_class_breakdown}
\end{figure}


We first decompose same-class substitution errors by articulatory feature differences for the strongest baseline model (PER = 14.46\%) \ref{fig:error_class_breakdown}). We analyze substitution errors since they constitue the majority of errors, indicating that model performance is primarily limited by confusion between acoustically similar phonemes rather than segmentation failures. The comprehensive breakdown is outlined in Appendix (Tables \ref{tab:appendix_model_errors_SDIC} and \ref{tab:appendix_model_errors_art}). For same-class substitutions, the errors exhibit highly structured patterns and predominantly involve low-dimensional feature changes. For consonants, nearly half of all substitutions differ only in voicing ($49.1\%$), while a further $31.7\%$ differ jointly in \textit{place} and \textit{manner}. Isolated place and manner errors are comparatively uncommon, indicating that consonant confusions are dominated by a small number of articulatory dimensions. Similarly, vowel substitutions are driven by \textit{height} differences ($50.9\%$), followed by combined \textit{height-backness} deviations. Overall, substitutions are concentrated along specific articulatory dimensions, with most confusions arising from differences in a single feature. This distribution is consistent with established phonetic properties of speech. Voicing is encoded through low-frequency periodicity that can be masked by noise, while vowel height is encoded through varying formant structure that can overlap across categories. Consequently, the baseline model captures coarse phonetic categories but fails to resolve fine-grained distinctions within them. These observations motivate our proposed MTL framework, which improves phoneme discrimination through articulatory supervision.

\begin{table}[!t]
\centering
\caption{Changes in recognition errors and articulatory feature substitutions between the best baseline (WavLM + Aug + MPL) and the best proposed model (HMTL + CA + Aug + MPL). $\Delta$ Domain Share is computed within consonant or vowel substitution classes and denotes normalized shift in error distribution.}
\label{tab:mtl_feature_changes}
\small
\begin{tabular}{llrr}
\toprule
Category & Metric & $\Delta$ (\%) & $\Delta$ Domain Share (\%)\\
\midrule
\multirow{1}{*}{Type}
& Cross-class & $+41.0$ & -\\
\midrule
\multirow{6}{*}{Articulatory}
& Place & $-3.24$ & $+1.01$\\
& Manner & $-3.86$ & $+0.66$\\
& Voicing & $-10.14$ & $-1.67$\\
& Height & $-5.65$ & $-0.06$\\
& Backness & $-5.83$ & $-0.07$\\
& Roundedness & $-4.73$ & $+0.14$\\
\bottomrule
\end{tabular}
\end{table}

To assess the impact of MTL, we examine substitution errors across articulatory feature dimensions (Table \ref{tab:mtl_feature_changes}). Relative to the strongest baselines, the proposed model reduces substitution errors across every feature, indicating more accurate articulatory discrimination under MTL. Reductions are concentrated along the most error-prone features, voicing and vowel height, which experience the some of the largest reductions ($-10.1\%$ and $-5.65\%$), respectively, whereas place and manner exhibit the smallest gains ($-3.24\%$ and $-3.86\%$). When normalized within each domain, voicing, the dominant source of consonant errors, decreases more rapidly than the remaining features, resulting in a redistribution of residual errors to place and manner. In contrast, in vowels, the relative distributions changes only marginally. Notably, we observe an increase in cross-class confusions under MTL. This can be be attributed to the lack of enforcement between the consonant-vowel boundary as within-class ambiguities are reduced, therefore increasing errors between /\texttt{R}/ and /\texttt{ER}/. These findings support the hypothesis that explicitly modeling articulatory features enables the model to disentangle correlated dimensions, thereby resolving the structural confusions observed in the baseline.

\subsection{Multi-Task-Learning Ablations}

\begin{figure}[ht]
    \centering
    \begin{subfigure}[t]{0.48\linewidth}
        \centering
        \includegraphics[width=\linewidth]{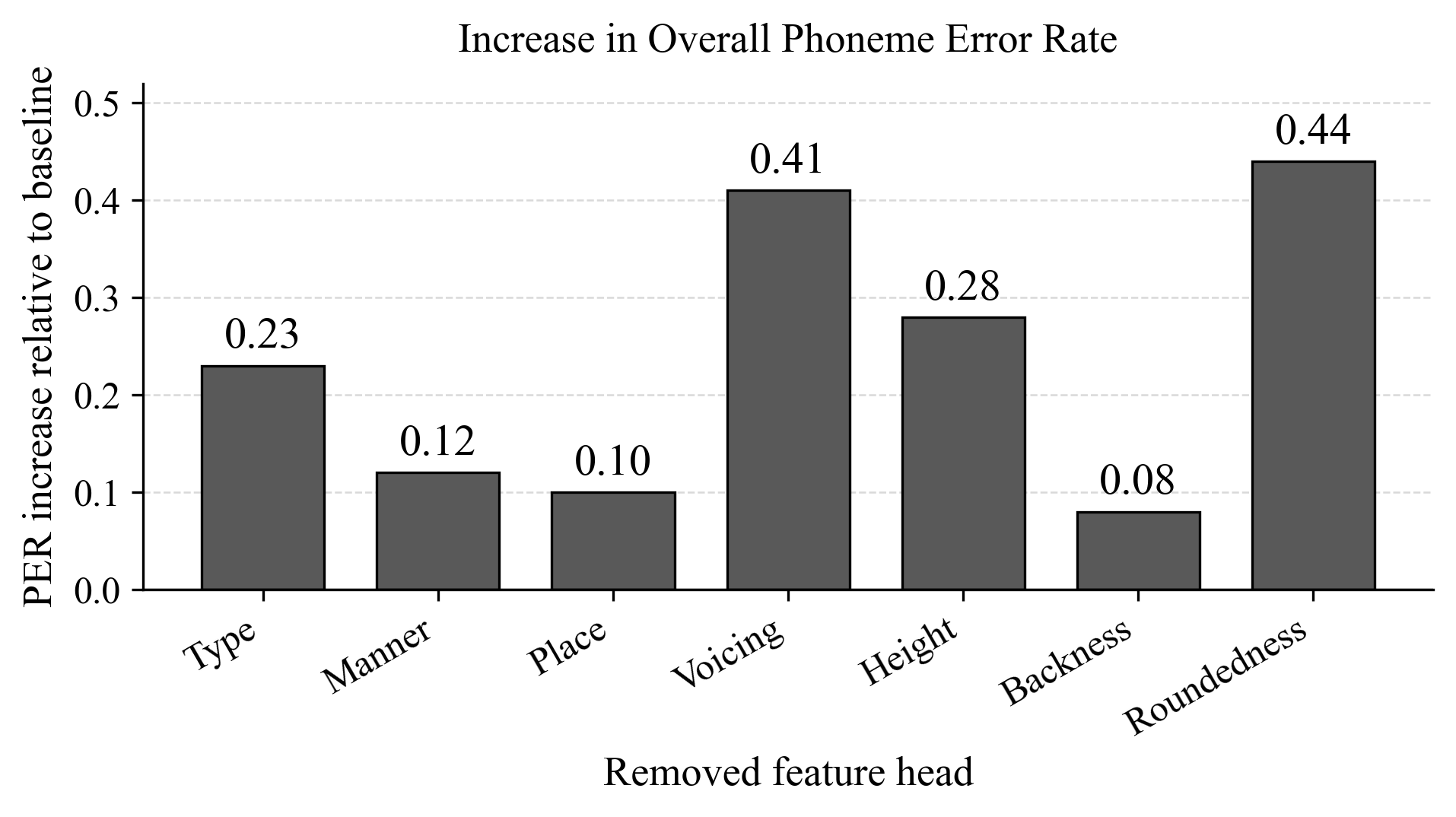}
        \caption{Aggregate PER increase by ablated head.}
        \label{fig:loo_ablation_bar}
    \end{subfigure}
    \hfill
    \begin{subfigure}[t]{0.48\linewidth}
        \centering
        \includegraphics[width=\linewidth]{ablation_delta_heatmap.png}
        \caption{Head-level PER deltas across features.}
        \label{fig:loo_ablation_heatmap}
    \end{subfigure}
    \caption{
Leave-one-out attention head ablation analysis of phoneme error rate (PER).
(a) Aggregate increase in PER when ablating heads associated with each phonological feature, showing the relative contribution of different task heads.
(b) Heatmap of PER changes for individual head removals (x-axis) across articulatory feature error types (y-axis), where color encodes the change in error relative to the baseline model.
}
\label{fig:loo_ablation}
\end{figure}

The results of the articulatory task head ablation study are summarized in Figure \ref{fig:loo_ablation}. The configuration with all heads achieved the lowest phoneme error rate (PER) of 14.13\%. Removing any individual articulatory head consistently degraded performance, with PER increases ranging from +0.08 to +0.44 relative to the baseline. The most substantial degradations were observed when removing the Roundedness (+0.44) and Voicing (+0.41) heads, followed by Height (+0.28) and Type (+0.23). In contrast, removing Backness (+0.08), Place (+0.10), and Manner (+0.12) resulted in comparatively smaller performance drops. These results indicate that not all articulatory features contribute equally, and that certain features, particularly Roundedness and Voicing, encode highly informative signals that are critical for accurate phoneme recognition.

A more fine-grained view of these effects is shown in the error-type breakdown heatmap (Figure \ref{fig:loo_ablation_heatmap}. Removing a feature head generally increases errors associated with that feature, demonstrably, removing roundedness significantly increases roundedness-related errors (+5.7\%) while also amplifying insertion (+2.9\%) and deletion errors (+2.5\%). These increased segmentation errors suggest that this feature plays a key role in stabilizing vowel boundaries. However, the effects extends beyond single features, revealing inter-dependencies between articulatory dimensions. Namely, removing the Voicing head leads to increases not only in voicing, but also in roundedness (+7.6\%), backness (+4.1\%), and insertions (+3.8\%), indicating cross-feature dependencies. In contrast, some removals lead to partial compensatory effects. To illustrate, removing Backness reduces deletion errors (-4.8\%) but increases insertion errors (+4.9\%), highlighting trade-offs in how the model distributes phonological information. 

From a linguistic perspective, features such as voicing rely on low-frequency periodicity that can easily be masked by noise 
\cite{stevens2000acoustic,ladefoged2014course}, making them inherently difficult to model, but highly valuable when learned, thereby creating substantial errors when removed. Height and Backness ablations may cause deletions to decrease as both are continuous, within-vowel features encoded through formant structure (F1 and F2), which varies smoothly across vowel realizations and provide weaker discriminative boundaries for phoneme presence detection. Removing these features reduces sensitivity to fine-grained vowel distinction, making the model less likely to reject ambiguous vowel segments and more prone to insertion and substitution errors. A similar logic applies to other localized blue cells: removing the Place and Voicing head slightly reduces manner-related errors, but omitting Manner yields small improvements in backness and deletions. Notably, some ablations lead to localized reductions in specific error types (blue regions), which may appear counterintuitive. For instance, removing Place or Voicing reduces manner related errors, while removing the Manner head improves backness and deletion errors. Place and manner are supported by partially redundant acoustic cues, including spectral shape, temporal structure, and formant transitions \cite{stevens2000acoustic,ladefoged2014course}, making these more robust to the removal of any single supervisory signal, resulting in even improved errors under ablation. However, these effects do not necessarily reflect direct linguistic similarity between features, but rather interactions within the learned representation that arise from ablations. Jointly learning features with competing representations can introduce representational competition, leading to localized improvements in some features when ablated, despite overall degradation in performance. The fact that removing any head worsens PER, while simultaneously redistributing feature-specific errors in structured ways, supports the claim that MTL improves performance not by uniform improvements, but by disentangling correlated articulatory dimensions according to linguistic reasoning.

\subsection{Augmentation Ablations} 
\begin{table}[h]
\centering
\caption{Leave-one-out ablation of audio augmentations.}
\label{tab:aug_ablation}
\begin{tabular}{lc}
\toprule
\textbf{Configuration} & \textbf{PER (\%)} \\
\midrule
All augmentations & \textbf{13.49} \\
$-$ Noise & 13.76 \\
$-$ VTLP & 13.80 \\
$-$ Phase perturbation & 13.83 \\
$-$ Prosody (Time-stretch) & 13.87 \\
$-$ Pitch shift & 13.94 \\
\bottomrule
\end{tabular}
\end{table}
To understand the contribution of different data augmentation strategies on model robustness, we performed a leave-one-out ablation study. We started with a baseline utilizing all augmentations (pitch shifting, time-stretching for prosody, phase perturbation, VTLP, and additive noise) and systematically removed one at a time. Overall, using our best-performing model configuration of HMTL+c.a.+MPL, our full augmentation suite achieves the lowest PER of 13.49\% (Table \ref{tab:aug_ablation}), demonstrating that a diverse set of regularization is strictly beneficial.

\textbf{Pitch and Prosody.}

Removing pitch shifting causes the most severe performance degradation, increasing PER to $13.94\%$. This aligns closely with the phonetic characteristics of L2 speech: non-native learners exhibit substantial fundamental frequency ($F_0$) variance, often inappropriately transferring L1 intonation or tone patterns to the L2 \cite{l2mennen}. Because phonemic contrasts in English rely heavily on the spectral envelope (formants) rather than absolute pitch, forcing the model to be invariant to $F_0$ prevents it from overfitting to irrelevant pitch contours. Similarly, removing prosodic time-stretching yields the second highest error ($13.87\%$). L2 learners are characterized by highly variable speaking rates, frequent hesitations, and inconsistent vowel elongation. Time-stretching naturally mimics these temporal distortions, improving the model's ability to locate phoneme boundaries despite dysfluent rhythms.

\textbf{Vocal Tract and Phase}

VTLP and phase perturbation address structural acoustic variations rather than behavioral ones. VTLP simulates speakers with differing physical vocal tract lengths, while phase perturbation simulates variations in microphone placement and room acoustics without fundamentally altering the linguistic magnitude spectrum. While less impactful than pitch and prosody, removing phase perturbation ($+0.34\%$ PER) and VTLP ($+0.31\%$ PER) still noticeably degrades performance, confirming that simulating diverse physical profiles and recording environments aids generalization safely. 

\textbf{Naturalness} 

Crucially, these augmentations mimic naturalistic variations better than purely synthetic perturbations (like masking or dropout). By restricting pitch shifts and time stretches within physically plausible bounds, we systematically expose the network to the specific axes of variance, pitch instability and temporal hesitation, that define the pathological and L2 speech distribution.

\section{Limitations}

Our approach introduces a structured framework for phoneme recognition using articulatory feature decomposition, but several limitations remain. The model operates at the phonemic level using a reduced ARPAbet inventory, which collapses fine-grained phonetic variation include allophonic distinctions such as aspiration, flapping, and vowel reduction, limiting the system’s ability to capture subtle but clinically relevant pronunciation errors. This is further compounded by ARPAbet itself, which omits several articulatory distinctions present in IPA, reducing the granularity of both supervision and evaluation. Although some phonemes posses multiple articulatory properties simultaneously, each auxiliary head is trained to predict a single discrete label. For example, /\texttt{W}/ exhibits both labial and dorsal articulations, yet the place classifier can predict only a single place category. This limits the model’s ability to capture complex articulatory structure. Furthermore, the proposed framework focuses exclusively on segmental phoneme recognition and does not model suprasegmental aspects of speech, including stress, rhythm, and intonation. As another limitation articulatory features are modeled as discrete categorical targets rather than continuous physiological processes. The framework therefore does not explicitly capture articulatory dynamics such as airflow, timing, or motion. Future work could explore richer inventories such as IPA that more faithfully represent these dynamics. Such extensions would allow the framework to better capture co-articulation, and the inherently continuous nature of speech production. Finally, an additional limitation of our work is that we evaluate only on L2-Arctic rather than a pathological dataset. We do this out of concerns for data pollution regarding existing large-scale pre-trained backbones being trained on most labeled phoneme and transcribed audio datasets, including older pathological speech datasets. L2-Arctic is relatively recent and has not, to our knowledge, been utilized as training data for the baselines we have chosen in this study. We thus motivate the research community for the creation of newer phoneme-labeled pathological speech datasets.

\section{Conclusion}

Pathological phoneme recognition remains challenging due to structured deviations in speech, limited data availability, and noisy or ambiguous labels. Conventional models, which treat phonemes as independent categories, are not well-equipped to capture these structured variations, particularly under data-constrained settings. In this work, we introduce an articulatory-feature-based architecture for phoneme recognition that explicitly incorporates physiological structure, and evaluate on L2 speech as a controlled proxy for systematic non-canonical and pathological phoneme variation. By modeling phoneme recognition as a composition of articulatory feature prediction tasks, integrating these representations through a cross-attention mechanism, and projecting into phoneme space after temporal modeling, our approach addresses structured errors that arise from independent phoneme modeling in standard baselines.

We demonstrate these improvements empirically through consistent reductions in phoneme error rate across strong baselines and ablated variants. Most noticeably the largest accuracy improvement was seen when combining our multi-task-learning framework with pseudo-labeling and strong augmentations, improving accuracy significantly compared to baseline methods even with the same data diversity techniques applied. This suggests that sufficient data diversity is necessary to fully realize the benefits of structured phoneme modeling.

In addition, we observe systematic reductions in structured error patterns. This is shown by an absolute decrease in errors and a redistribution away from within-feature confusions toward cross-feature errors. When analyzing error reduction across features, the addition of MTL affects the particular features that contributed most to errors observed in the baseline. Ablation studies further show that removing individual articulatory feature heads leads to predictable, linguistically interpretable degradation in performance. Vowel features in particular prove to be highly impactful; we find that including the vowel roundedness task head significantly reduces associated roundedness recognition error rates, and that the consonant voicing head further helps this category of confusions. While voicing and roundedness are  distinct features, they may interact indirectly in the acoustic signal through co-articulation or overlap and other learned statistical regularities. We therefore interpret their joint importance in the model as reflecting correlated acoustic cues rather than a direct linguistic dependency.

Overall, our results highlight the importance of incorporating articulatory structure into speech recognition systems, particularly for non-canonical and clinical speech settings. While our experiments use accented speech as a proxy, future work should extend this framework to clinically validated pathological datasets and explore richer articulatory representations, including multi-label and continuous formulations. Improving pathological phoneme recognition has the potential to support digital clinical tools and enhance feedback in speech-language pathology workflows, motivating future attention in this field.

\section{References}
\bibliographystyle{compling}
\bibliography{refs}

@article{yang2022,
  title={Improving mispronunciation detection with wav2vec2-based momentum pseudo-labeling for accentedness and intelligibility assessment},
  author={Yang, Mu and Hirschi, Kevin and Looney, Stephen D and Kang, Okim and Hansen, John HL},
  journal={arXiv preprint arXiv:2203.15937},
  year={2022}
}

@misc{asha2023poll,
  author = {{American Speech-Language-Hearing Association}},
  title = {Poll Shows Increases in Hearing, Speech, and Language Referrals, More Communication Challenges in Young Children},
  year         = {2023},
  url = {https://www.asha.org/news/2023/poll-shows-increases-in-hearing-speech-and-language-referrals-more-communication-challenges-in-young-children/}}

@article{cohen2025,
    title = {Robust prosody modeling for synthetic speech detection},
    journal = {Speech Communication},
    volume = {174},
    pages = {103283},
    year = {2025},
    issn = {0167-6393},
    doi = {10.1016/j.specom.2025.103283},
    url = {https://www.sciencedirect.com/science/article/pii/S0167639325000986},
    author = {Ariel Cohen and Denis Shyrman and Aleksandr Solonskyi and Roman Frenkel and Arkady Krishtul and Oren Gal},
    }

@article{strombergsson2020,
title = {Audience Response System-Based Evaluation of Intelligibility of Children’s Connected Speech – Validity, Reliability and Listener Differences},
journal = {Journal of Communication Disorders},
volume = {87},
pages = {106037},
year = {2020},
issn = {0021-9924},
doi = {10.1016/j.jcomdis.2020.106037},
url = {https://www.sciencedirect.com/science/article/pii/S0021992420301052},
author = {Sofia Strömbergsson and Katarina Holm and Jens Edlund and Tove Lagerberg and Anita McAllister},
}

@article{berisha2024clinicalai,
  title={Responsible development of clinical speech AI: Bridging the gap between clinical research and technology},
  author={Berisha, Visar and Liss, Julie M.},
  journal={npj Digital Medicine},
  volume={7},
  pages={208},
  year={2024},
  doi={10.1038/s41746-024-01199-1}
}

@misc{clevelandclinic2025dysarthria,
  author       = {{Cleveland Clinic}},
  title        = {Dysarthria (Slurred Speech): Symptoms, Causes \& Treatment},
  year         = {2025},
  url          = {https://my.clevelandclinic.org/health/diseases/17653-dysarthria},
  note         = {Medically reviewed; Accessed: 2026-04-11}
}

@BOOK{duffy2019,
  title     = "Motor Speech Disorders: Substrates, Differential Diagnosis, and Management",
  author    = "Duffy, Joseph R",
  publisher = "Mosby",
  edition   =  4,
  month     =  dec,
  year      =  2019,
  address   = "St. Louis, MO",
  language  = "en"
}

@article{chen2022wavlm,
   title={WavLM: Large-Scale Self-Supervised Pre-Training for Full Stack Speech Processing},
   volume={16},
   ISSN={1941-0484},
   url={http://dx.doi.org/10.1109/JSTSP.2022.3188113},
   DOI={10.1109/jstsp.2022.3188113},
   number={6},
   journal={IEEE Journal of Selected Topics in Signal Processing},
   publisher={Institute of Electrical and Electronics Engineers (IEEE)},
   author={Chen, Sanyuan and Wang, Chengyi and Chen, Zhengyang and Wu, Yu and Liu, Shujie and Chen, Zhuo and Li, Jinyu and Kanda, Naoyuki and Yoshioka, Takuya and Xiao, Xiong and Wu, Jian and Zhou, Long and Ren, Shuo and Qian, Yanmin and Qian, Yao and Wu, Jian and Zeng, Michael and Yu, Xiangzhan and Wei, Furu},
   year={2022},
   month=oct, pages={1505–1518} 
   }

@article{baevski2020wav2vec2,
  title={wav2vec 2.0: A framework for self-supervised learning of speech representations},
  author={Baevski, Alexei and Zhou, Yuhao and Mohamed, Abdelrahman and Auli, Michael},
  journal={Advances in neural information processing systems},
  volume={33},
  pages={12449--12460},
  year={2020}
}

@article{torgo2012,
author = {Rudzicz, Frank and Namasivayam, Aravind Kumar and Wolff, Talya},
title = {The TORGO database of acoustic and articulatory speech from speakers with dysarthria},
year = {2012},
issue_date = {December  2012},
publisher = {Springer-Verlag},
address = {Berlin, Heidelberg},
volume = {46},
number = {4},
issn = {1574-020X},
url = {https://doi.org/10.1007/s10579-011-9145-0},
doi = {10.1007/s10579-011-9145-0},
journal = {Lang. Resour. Eval.},
month = dec,
pages = {523–541},
numpages = {19}
}

@article{FATEHI2025103151,
title = {An overview of high-resource automatic speech recognition methods and their empirical evaluation in low-resource environments},
journal = {Speech Communication},
volume = {167},
pages = {103151},
year = {2025},
issn = {0167-6393},
doi = {doi.org/10.1016/j.specom.2024.103151},
url = {https://www.sciencedirect.com/science/article/pii/S0167639324001225},
author = {Kavan Fatehi and Mercedes {Torres Torres} and Ayse Kucukyilmaz}
}

@article{HOSOM2009352,
title = {Speaker-independent phoneme alignment using transition-dependent states},
journal = {Speech Communication},
volume = {51},
number = {4},
pages = {352-368},
year = {2009},
issn = {0167-6393},
doi = {10.1016/j.specom.2008.11.003},
url = {https://www.sciencedirect.com/science/article/pii/S0167639308001775},
author = {John-Paul Hosom}
}

@inproceedings{l2arctic,
    author={Guanlong {Zhao} and Sinem {Sonsaat} and Alif {Silpachai} and Ivana {Lucic} and Evgeny {Chukharev-Hudilainen} and John {Levis} and Ricardo {Gutierrez-Osuna}},
    title={L2-ARCTIC: A Non-native English Speech Corpus},
    year={2018},
    booktitle={Interspeech 2018},
    pages={2783–2787},
    doi={10.21437/Interspeech.2018-1110},
    url={http://dx.doi.org/10.21437/Interspeech.2018-1110}
}

@article{FARISH2020100910,
title = {Listener perceptions of foreignness, precision, and accent attribution in a case of foreign accent syndrome},
journal = {Journal of Neurolinguistics},
volume = {55},
pages = {100910},
year = {2020},
issn = {0911-6044},
doi = {10.1016/j.jneuroling.2020.100910},
url = {https://www.sciencedirect.com/science/article/pii/S0911604419301319},
author = {Brian A. Farish and Lori A. Davis and Laura D. Wilson}
}

@incollection{flege,
  author    = {Flege, James Emil},
  title     = {Second language speech learning: Theory, findings and problems},
  booktitle = {Speech perception and linguistic experience: Issues in cross-language research},
  editor    = {Strange, Winifred},
  pages     = {233-277},
  year      = {1995},
  publisher = {York Press},
  address   = {Baltimore}
}

@incollection{besttyler,
  author    = {Best, Catherine T. and Tyler, Michael D.},
  title     = {Nonnative and second-language speech perception: Commonalities and complementarities},
  booktitle = {Second language speech learning: The role of language experience in speech perception and production},
  editor    = {Munro, Murray J. and Bohn, Ocke-Schwen},
  pages     = {13-34},
  year      = {2007},
  publisher = {John Benjamins},
  address   = {Amsterdam}
}

@ARTICLE{Kent2003-ya,
  title     = "Toward an acoustic typology of motor speech disorders",
  author    = "Kent, Ray D and Kim, Y J",
  journal   = "Clin. Linguist. Phon.",
  publisher = "Informa UK Limited",
  volume    =  17,
  number    =  6,
  pages     = "427--445",
  month     =  sep,
  year      =  2003,
  language  = "en"
}

@inproceedings{jaitly2013vocal,
  title={Vocal tract length perturbation (VTLP) improves speech recognition},
  author={Jaitly, Navdeep and Hinton, Geoffrey E},
  booktitle={Proc. ICML workshop on deep learning for audio, speech and language},
  volume={117},
  pages={21},
  year={2013}
}

@inproceedings{chengxilei,
author = {Lei, Chengxi and Singh, Satwinder and Hou, Feng and Jia, Xiaoyun and Wang, Ruili},
title = {PhasePerturbation: Speech Data Augmentation via Phase Perturbation for Automatic Speech Recognition},
year = {2023},
isbn = {9798400703263},
publisher = {Association for Computing Machinery},
address = {New York, NY, USA},
url = {https://doi.org/10.1145/3611380.3628555},
doi = {10.1145/3611380.3628555},
booktitle = {Proceedings of the 5th ACM International Conference on Multimedia in Asia Workshops},
articleno = {2},
numpages = {6},
location = {Tainan, Taiwan},
series = {MMAsia '23 Workshops}
}

@book{ladefoged2014course,
  author    = {Ladefoged, Peter and Johnson, Keith},
  title     = {A Course in Phonetics},
  edition   = {6},
  year      = {2014},
  publisher = {Cengage Learning}
}

@inproceedings{stouten2006speech,
  title     = {{Speech recognition with phonological features: some issues to attend}},
  author    = {Frederik Stouten and Jean-Pierre Martens},
  year      = {2006},
  booktitle = {{Interspeech 2006}},
  pages     = {paper 1081-Mon2BuP.4},
  doi       = {10.21437/Interspeech.2006-121},
  issn      = {2958-1796},
}

@inproceedings{Thienpondt_2025,
   title={Weakly Supervised Phonological Features for Pathological Speech Analysis},
   url={http://dx.doi.org/10.1109/ICASSP49660.2025.10888038},
   DOI={10.1109/icassp49660.2025.10888038},
   booktitle={ICASSP 2025 - 2025 IEEE International Conference on Acoustics, Speech and Signal Processing (ICASSP)},
   publisher={IEEE},
   author={Thienpondt, Jenthe and Vanderreydt, Geoffroy and Hammami, Abdessalem and Demuynck, Kris},
   year={2025},
   month=apr, pages={1–5} }

@inproceedings{tadavarthy2024phonological,
  title     = {{Phonological Feature Detection for US English using the Phonet Library}},
  author    = {Harsha Veena Tadavarthy and Austin Jones and Margaret E. L. Renwick},
  year      = {2024},
  booktitle = {{Interspeech 2024}},
  pages     = {1515--1519},
  doi       = {10.21437/Interspeech.2024-318},
  issn      = {2958-1796},
}

@inproceedings{xu2022simple,
  author    = {Xu, Qiantong and Baevski, Alexei and Auli, Michael},
  title     = {Simple and Effective Zero-shot Cross-lingual Phoneme Recognition},
  booktitle = {Interspeech 2022},
  year      = {2022},
  publisher = {ISCA},
  doi = {10.21437/Interspeech.2022-60}
}

@article{SHAHIN2025103249,
title = {Phonological level wav2vec2-based Mispronunciation Detection and Diagnosis method},
journal = {Speech Communication},
volume = {173},
pages = {103249},
year = {2025},
issn = {0167-6393},
doi = {10.1016/j.specom.2025.103249},
url = {https://www.sciencedirect.com/science/article/pii/S0167639325000640},
author = {Mostafa Shahin and Julien Epps and Beena Ahmed}
}

@inproceedings{PanPhonMortensen-et-al:2016,
  author    = {David R. Mortensen and
               Patrick Littell and
               Akash Bharadwaj and
               Kartik Goyal and
               Chris Dyer and
               Lori S. Levin},
  title     = {PanPhon: {A} Resource for Mapping {IPA} Segments to Articulatory Feature Vectors},
  booktitle = {Proceedings of {COLING} 2016, the 26th International Conference on Computational Linguistics: Technical Papers},
  pages     = {3475--3484},
  publisher = {{ACL}},
  year      = {2016}
}

@book{Glocker2024HMTL,
author="Glocker, Kevin
and Georges, Munir",
title="Hierarchical Multi-task Learning with Articulatory Attributes for Cross-Lingual Phoneme Recognition",
bookTitle="Practical Solutions for Diverse Real-World NLP Applications",
year="2024",
publisher="Springer International Publishing",
address="Cham",
pages="59--75",
isbn="978-3-031-44260-5",
doi="10.1007/978-3-031-44260-5_4",
url="https://doi.org/10.1007/978-3-031-44260-5_4"
}

@inproceedings{elkheir2023multiview,
  title     = {{Multi-View Multi-Task Representation Learning for Mispronunciation Detection}},
  author    = {Yassine {EL Kheir} and Shammur Chowdhury and Ahmed Ali},
  year      = {2023},
  booktitle = {{9th Workshop on Speech and Language Technology in Education (SLaTE)}},
  pages     = {86--90},
  doi       = {10.21437/SLaTE.2023-18},
  issn      = {2311-4975},
}

@article{PANG2026131930,
title = {A multi-task hierarchical deep reinforcement network approach for word level pronunciation assessment},
journal = {Expert Systems with Applications},
volume = {317},
pages = {131930},
year = {2026},
issn = {0957-4174},
doi = {10.1016/j.eswa.2026.131930},
url = {https://www.sciencedirect.com/science/article/pii/S0957417426008432},
author = {Xudong Pang and Aishan Wumaier and Wenwen Lu and Silajiaihemaiti Ruzemaimaiti and Ligong Lei},
}

@inproceedings{librispeech,
  author={Panayotov, Vassil and Chen, Guoguo and Povey, Daniel and Khudanpur, Sanjeev},
  booktitle={2015 IEEE International Conference on Acoustics, Speech and Signal Processing (ICASSP)}, 
  title={Librispeech: An ASR corpus based on public domain audio books}, 
  year={2015},
  volume={},
  number={},
  pages={5206-5210},
  doi={10.1109/ICASSP.2015.7178964}
  }

@book{ipa,
  author    = {{International Phonetic Association}},
  title     = {Handbook of the International Phonetic Association: A Guide to the Use of the International Phonetic Alphabet},
  year      = {1999},
  publisher = {Cambridge University Press}
}

@misc{arpabet,
  author       = {{Carnegie Mellon University}},
  title        = {The CMU Pronouncing Dictionary},
  year         = {1998},
  howpublished = {\url{http://www.speech.cs.cmu.edu/cgi-bin/cmudict}},
  note         = {Defines ARPAbet phoneme set}
}

@INPROCEEDINGS{li2020allosaurus,
  author={Li, Xinjian and Dalmia, Siddharth and Li, Juncheng and Lee, Matthew and Littell, Patrick and Yao, Jiali and Anastasopoulos, Antonios and Mortensen, David R. and Neubig, Graham and Black, Alan W and Metze, Florian},
  booktitle={ICASSP 2020 - 2020 IEEE International Conference on Acoustics, Speech and Signal Processing (ICASSP)}, 
  title={Universal Phone Recognition with a Multilingual Allophone System}, 
  year={2020},
  volume={},
  number={},
  pages={8249-8253},
  doi={10.1109/ICASSP40776.2020.9054362}}

@article{ravanelli2021speechbrain,
  title={SpeechBrain: A general-purpose speech toolkit},
  author={Mirco Ravanelli and Titouan Parcollet and Peter Plantinga and Aku Rouhe and Samuele Cornell and Loren Lugosch and Cem Subakan and Nauman Dawalatabad and Abdelwahab Heba and Jianyuan Zhong and Ju-Chieh Chou and Sung-Lin Yeh and Szu-Wei Fu and Chien-Feng Liao and Elena Rastorgueva and François Grondin and William Aris and Hwidong Na and Yan Gao and Renato De Mori and Yoshua Bengio},
  journal={arXiv preprint arXiv:2106.04624},
  year={2021}
}

@inproceedings{uaspeech,
  title     = {{Dysarthric speech database for universal access research}},
  author    = {Heejin Kim and Mark Hasegawa-Johnson and Adrienne Perlman and Jon Gunderson and Thomas S. Huang and Kenneth Watkin and Simone Frame},
  year      = {2008},
  booktitle = {Interspeech 2008},
  pages     = {1741--1744},
  doi       = {10.21437/Interspeech.2008-480},
  issn      = {2958-1796},
}

@inproceedings{nemours,
  title     = {{The nemours database of dysarthric speech}},
  author    = {Xavier Menéndez-Pidal and James B. Polikoff and Shirley M. Peters and Jennie E. Leonzio and H. T. Bunnell},
  year      = {1996},
  booktitle = {{4th International Conference on Spoken Language Processing (ICSLP 1996)}},
  pages     = {1962--1965},
  doi       = {10.21437/ICSLP.1996-503},
  issn      = {2958-1796},
}

@inproceedings{ctcgraves,
author = {Graves, Alex and Fern\'{a}ndez, Santiago and Gomez, Faustino and Schmidhuber, J\"{u}rgen},
title = {Connectionist temporal classification: labelling unsegmented sequence data with recurrent neural networks},
year = {2006},
isbn = {1595933832},
publisher = {Association for Computing Machinery},
address = {New York, NY, USA},
url = {https://doi.org/10.1145/1143844.1143891},
doi = {10.1145/1143844.1143891},
booktitle = {Proceedings of the 23rd International Conference on Machine Learning},
pages = {369–376},
numpages = {8},
location = {Pittsburgh, Pennsylvania, USA},
series = {ICML '06}
}

@article{bilstmgraves,
title = {Framewise phoneme classification with bidirectional LSTM and other neural network architectures},
journal = {Neural Networks},
volume = {18},
number = {5},
pages = {602-610},
year = {2005},
issn = {0893-6080},
doi = {10.1016/j.neunet.2005.06.042},
url = {https://www.sciencedirect.com/science/article/pii/S0893608005001206},
author = {Alex Graves and Jürgen Schmidhuber}
}

@inproceedings{ozansener,
author = {Sener, Ozan and Koltun, Vladlen},
title = {Multi-task learning as multi-objective optimization},
year = {2018},
publisher = {Curran Associates Inc.},
address = {Red Hook, NY, USA},
booktitle = {Proceedings of the 32nd International Conference on Neural Information Processing Systems},
pages = {525–536},
numpages = {12},
location = {Montr\'{e}al, Canada},
series = {NIPS'18}
}

@article{mgda,
title = {Multiple-gradient descent algorithm (MGDA) for multiobjective optimization},
journal = {Comptes Rendus Mathematique},
volume = {350},
number = {5},
pages = {313-318},
year = {2012},
issn = {1631-073X},
doi = {10.1016/j.crma.2012.03.014},
author = {Jean-Antoine Désidéri}
}

@inproceedings{park19e_interspeech,
  title     = {{SpecAugment: A Simple Data Augmentation Method for Automatic Speech Recognition}},
  author    = {Daniel S. Park and William Chan and Yu Zhang and Chung-Cheng Chiu and Barret Zoph and Ekin D. Cubuk and Quoc V. Le},
  year      = {2019},
  booktitle = {{Interspeech 2019}},
  pages     = {2613--2617},
  doi       = {10.21437/Interspeech.2019-2680},
  issn      = {2958-1796},
}

@article{praat,
author = {Boersma, Paul and Weenink, David},
year = {2001},
month = {01},
pages = {341-345},
title = {PRAAT, a system for doing phonetics by computer},
volume = {5},
journal = {Glot international}
}

@inproceedings{ryu12_interspeech,
  title     = {{Comparing transcription agreement on non-native English speech corpus between native and non-native annotators}},
  author    = {Hyuksu Ryu and Sunhee Kim and Minhwa Chung},
  year      = {2012},
  booktitle = {{Interspeech 2012}},
  pages     = {2366--2369},
  doi       = {10.21437/Interspeech.2012-620},
  issn      = {2958-1796},
}

@book{stevens2000acoustic,
  author    = {Stevens, Kenneth N.},
  title     = {Acoustic Phonetics},
  year      = {2000},
  publisher = {MIT Press},
  doi = {10.7551/mitpress/1072.001.0001}
}

@book{l2mennen,
author="Mennen, Ineke",
title="Beyond Segments: Towards a L2 Intonation Learning Theory",
bookTitle="Prosody and Language in Contact: L2 Acquisition, Attrition and Languages in Multilingual Situations",
year="2015",
publisher="Springer Berlin Heidelberg",
address="Berlin, Heidelberg",
pages="171--188",
isbn="978-3-662-45168-7",
doi="10.1007/978-3-662-45168-7_9",
}

\newpage
\appendix

\appendixsection{Additional Inadequate IPA-ARPAbet Articulatory Mappings}

\begin{table}[h]
\caption{Language-Specific missing or ambiguous mappings between ARPAbet, IPA, and L2-ARCTIC annotations.}
\centering
\begin{tabular}{p{0.05\textwidth}lp{0.12\textwidth}lll}
\hline
\textbf{ARPA} & \textbf{IPA} & \textbf{L2-Arctic \newline Substitute} & \textbf{Example} & \textbf{Comments} &\\
\hline
- & \textipa{\textcrh} & HH & \textipa{\textcrh am"ma:m } & Arabic pharyngeal &  \\
-   & \textipa{\textrtails} & SH & \textipa{\textrtails\textrhookschwa} & Mandarin retroflex fricative  \\
-   & \textipa{k\super h} & K HH & \textipa{k\super ha:na:} & Hindi aspiration\\
-   & \textipa{J} & Y & \textipa{kam\textctj i} & Korean nasal \\
-   & \textipa{B} & B & \textipa{aBa} & Spanish bilabial fricative \\
Q  & \textipa{\textglotstop} & - & \textipa{ba\textglotstop} & Vietnamese glottal \\
\hline
\end{tabular}
\label{tab:lang_arpa_ipa_missing}
\end{table}

\begin{table}[ht!]
\caption{Diphthongs and IPA symbols with diacritics}
\centering
\begin{tabular}{llp{0.15\textwidth}l}
\hline
\textbf{Index} & \textbf{ARPA} & \textbf{L2-ARCTIC IPA} & \textbf{Comments}\\
\hline
5 & AW & \textipa{aU} & diphthong\\
7 & AY & \textipa{aI} & diphthong \\
9 & CH & \textipa{tS} & affricate \\
13 & ER & \textipa{\textrhookrevepsilon} & rhotic allophone of \textipa{3}\\
14 & EY & \textipa{eI} & diphthong \\
20 & JH & \textipa{dZ} & affricate \\
22 & L & \textipa{\textltilde} & velarized allophone of \textipa{l}\\
26 & OW & \textipa{oU} & diphthong \\
27 & OY & \textipa{OI} & diphthong\\
\hline
\end{tabular}
\label{tab:arpa_ipa_dipthong}
\end{table}

\renewcommand{\arraystretch}{0.8}
\begin{table}[ht!]
\caption{Phonetic distinctions collapsed by the L2-ARCTIC ARPAbet inventory.}
\centering
\begin{tabular}{p{0.05\textwidth}lp{0.12\textwidth}lll}
\hline
\textbf{ARPA} & \textbf{IPA} & \textbf{L2-Arctic \newline Substitute} & \textbf{Example} & \textbf{Comments} &\\
\hline
DX  & \textipa{\textfishhookr} & T  & bottle /\textipa{"bA\textfishhookr \s{l}}/ & approximation & \\
EL  & \textipa{\s{l}} & AH L & bottle /\textipa{"bA\textfishhookr \s{l}}/ & closure & \\
EM  & \textipa{\s{m}} & AH M & rhythm /\textipa{"rID\textsyllabic m}/  & closure &\\ 
EN  & \textipa{\s{n}} & AH N & button /\textipa{"bVt\textsyllabic n}/ & closure &\\ 
NX  & \textipa{\~n} & N & winner /\textipa{"wI\~n@r}/ & nasalized &\\
WH  & \textipa{\textturnw} & W & why /\textipa{\textturnw aI}/ & approximation &\\
AXR & \textipa{\textrhookschwa} & ER  & forward /\textipa{"fOrw\textrhookschwa d}/& rhotic approximation &\\
IX  & \textipa{1} & IH & rabbit /\textipa{"r\ae b1t}/ & approximation &\\
UX  & \textipa{\textbaru} & UH & dude /\textipa{d\textbaru d}/& approximation &\\
\hline
\end{tabular}
\label{tab:arpa_ipa_missing}
\end{table}
\renewcommand{\arraystretch}{1.0}

\appendixsection{Additional Evaluations and Results}
We detail PERs for several other ablation combinations of architectures, training strategy, and augmentation presence below.

\begin{table}[h]
\caption{Change in errors (Best baseline to best model).}

\centering
\small
\begin{tabular}{lrrr}
\toprule
Metric & Best Baseline & Best HMTL & $\Delta\%$ \\
\midrule
PER ($\%$)                     & 14.46               &   13.47       & $-6.8\%$ \\
Substitutions (within-class)   &        2889         &          2732 & $-5.4\%  $   \\     
Deletions                      &     705             &       520     & $   -26.2\%     $    \\ 
Insertions                     &     495             &       501     & $   +1.2\%     $ \\
Cross-class subs.              &     117             &       165     & $+41.0\%$ \\
\bottomrule
\end{tabular}
\label{tab:model_errors}
\end{table}

\begin{table}[h]
\caption{Phoneme error types by model.}

\centering
\small
\begin{tabular}{lrrrr}
\toprule
Model & Substitutions & Deletions & Insertions & Cross-class \\
\midrule
Wav2Vec2Phoneme       & 3143 &  523  &  642   &   122  \\
WavLM-base-plus    & 3003 &  776  &  499   &   121  \\
WavLM + Aug                       & 3055 &  673  &  555   &   123  \\
WavLM + MPL                       & 3067 &  657  &  504   &   123  \\
\textbf{WavLM + Aug. + MPL} &   3006    &    705   &    495   &   117    \\
\midrule
Parallel MTL & 2978 &  381 &  745 &  207 \\
Parallel MTL + Aug. & 2914 &  414 &  715 &  190 \\
Parallel MTL + Aug.\ + MPL & 2871 &  365 &  766 &  207 \\
\midrule
Hierarchical MTL               & 2887 &  567 &  491 &  196 \\
Hierarchical MTL + Aug.        & 2860 &  457 &  542 &  179 \\
Hierarchical MTL + Aug.\ + MPL & 2793 &  534 &  478 &  173 \\
\midrule
Hierarchical MTL + CA               & 2746 &  542 &  494 &  173 \\
Hierarchical MTL + CA + Aug.        & 2699 &  576 &  488 &  169 \\
\textbf{Hierarchical MTL + CA + Aug.\ + MPL} & 2732 &  520 &  501 &  165 \\
\bottomrule
\end{tabular}
\label{tab:appendix_model_errors_SDIC}
\end{table}

\begin{table}[h]
\caption{Phoneme feature errors by model.}

\centering
\small
\setlength{\tabcolsep}{3pt} 
\begin{tabular}{lrrrrrrr}
\toprule
Model & Cross-class & Place & Manner & Voicing & Height & Backness & Roundedness \\
\midrule
Wav2Vec2Phoneme              &    122  &  716  &   603  &    810   &   1105  &     431   &  286\\
WavLM-base-plus              &    121  &  652  &   549  &    769   &   1094  &     441   &  275\\
WavLM + Aug                  &    123  &  665  &   565  &    794   &   1087  &     444   &  290\\
WavLM + MPL                  &    123  &  640  &   540  &    830   &   1074  &     436   &  301\\
\textbf{WavLM + Aug + MPL}            &    117  &  679  &   570  &    789   &   1044  &     429   &  275\\
\midrule
Parallel MTL                 &    207  &  715  &   627  &   786    &   1067  &     445   &  298\\
Parallel MTL+Aug.            &    190  &  690  &   609  &   771    &   1042  &     440   &  291\\
Parallel MTL+Aug+MPL         &    207  &  686  &   602  &   769    &   1021  &     429   &  283\\
\midrule
Hierarchical MTL             &    196  &  669  &   576  &   798    &   1048  &     412   &  269\\
Hierarchical MTL+Aug.        &    179  &  672  &   578  &   734    &   1062  &     420   &  264\\
Hier MTL+Aug+MPL             &    173  &  666  &   555  &   717    &   1017  &     415   &  282\\
\midrule
Hier MTL+CA ~                &    173  &  646  &   551  &   743    &    988  &     411   &  266\\
Hier MTL+CA+Aug.             &    169  &  648  &   538  &   709    &    976  &     411   &  252\\
\textbf{Hier MTL+CA+Aug+MPL}        &    165  &  657  &   548  &   709    &    985  &     404   &  262\\
\bottomrule
\end{tabular}
\label{tab:appendix_model_errors_art}
\end{table}

\end{document}